\documentclass{raa}           
\usepackage{graphicx,times}
\usepackage{threeparttable, tablefootnote}
\usepackage{natbib}
\usepackage{amssymb,amsmath}
\bibpunct{(}{)}{;}{a}{}{,}
\usepackage[a4paper=true,dvipdfm=true,pagebackref=true]{hyperref}
\hypersetup{pdftitle = The title of my PDF, pdfauthor = My name, pdfsubject= The subject, pdfkeywords = keyword1 keyword2 keyword3} 
\hypersetup{colorlinks = true, linkcolor = green, anchorcolor = red, citecolor = blue, filecolor = red, pagecolor = red, urlcolor = red}

\begin{document}

   \title{Co-evolution of Nuclear Rings, Bars and the Central Intensity Ratio of their Host Galaxies}

 \volnopage{ {\bf 20XX} Vol.\ {\bf X} No. {\bf XX}, 000--000}
   \setcounter{page}{1}

   \author{S. Aswathy\inst{}, C. D. Ravikumar\inst{}
   }

   \institute{ Department of Physics, University of Calicut, Malappuram 673635, 
India; {\it aswathysahaj@gmail.com}\\
\vs \no
   {\small Received 20XX Month Day; accepted 20XX Month Day}
}

\abstract{Using a sample of 13 early-type spiral galaxies hosting nuclear rings, we report remarkable correlations between the properties of the nuclear rings and the central intensity ratio (CIR) of their host galaxies. The CIR, a function of intensity of light within the central 1.5 and 3 arcsec region, is found to be a vital parameter in galaxy evolution, as it shares strong correlations with many structural and dynamical properties of early-type galaxies, including mass of the central supermassive black hole (SMBH). We use archival \textsc{$HST$} images for aperture photometry at the centre of the galaxy image to compute the CIR. We observe that the relative sizes of nuclear rings and ring cluster surface densities strongly correlate with the CIR. These correlations suggest reduced star formation in the centres of galaxies hosting small and dense nuclear rings. This scenario appears to be a consequence of strong bars as advocated by the significant connection observed between the CIR and bar strengths. In addition, we observe that the CIR is closely related with the integrated properties of the stellar population in the nuclear rings associating the rings hosting older and less massive star clusters with low values of CIR. Thus, the CIR can serve as a crucial parameter in unfolding the coupled evolution of bars and rings as it is intimately connected with both their properties. 
}

   \authorrunning{S. Aswathy $\&$  C. D. Ravikumar}         
   \titlerunning{Co-evolution of nuclear rings}  
   \maketitle
\section{INTRODUCTION}          
\label{sect:intro}
Nuclear rings, also known as the circum-nuclear starburst rings, preferentially reside in barred spirals that constitute nearly two-thirds of normal spirals in  the local Universe \citep{KN99,LA04}. These rings with their intense star formation activities and close association with their host galaxy's structural parameters are believed to be ideal laboratories for understanding the secular evolution of their host galaxies \citep{MA95,BC96,KK04,MA08,KN15}. 

The origin of nuclear rings is thought to be the gravitational torque formed as a result of the non-axisymmetric perturbations emanating from the bars, spiral arms or ovals \citep{SH90,AT94,CO01}. The molecular gas, driven by shocks, flows inwards along the dust lanes on the leading edge of the bar and loses its angular momentum thereby spiraling into the circum-nuclear region \citep{KK04,KN05}. This inflow of matter can cause a burst of star formation activities and the matter gets trapped by the resonances in inner stellar orbits known as Inner Lindblad Resonances  (ILRs, \citealt{CG85,SB90,AT94,KNB95,KNA95,BC96}). There are various other theories about the origin of the rings such as \citet{SB90} which predicts  influence of  axisymmetric bulge in ring formation. Yet another explanation suggests  the nuclear rings as remnants of the nuclear starbursts resulting from the high surface densities of gas in the central region \citep{KE93}. Many simulations have been carried out based on various theories over the years \citep[see.g.][]{CG85,AT92,PI95}.

The nuclear rings are thus believed to be tracers of star formation in such galaxies. These rings are linked with the formation of massive clusters of stars near the centres of galaxies including Young Massive Clusters (YMCs)  \citep{MB01,DM17}. These clusters can also be used to constrain properties of their  host galaxies \citep{WS99,LI15}. Recent studies indicate that stronger bars have lower star formation at their central regions \citep{KM17,CM18}. In this light, we perform an optical study at the centres of nearby early-type spirals hosting nuclear ring and its associated clusters using a sample of 13 galaxies. We use a newly introduced parameter known as the central intensity ratio (CIR, \citealt{AR18}) to probe the interplay between the bars and nuclear rings in the secular evolution process. The CIR is the ratio of intensity of light contained in a central circular aperture of radius \textit{r} to that of an outer shell of width \textit{r}.

The CIR for early-type galaxies is reported to be anti-correlated with the mass of the supermassive black holes (SMBH,  \citealt{AR18}). It is closely related to various structural and dynamical properties of host galaxies. The CIR is also found to contain information about the star formation near the central region of these galaxies. Thus, the CIR can serve as an ideal tool in studying the star forming nuclear rings. Since majority of our sample galaxies are barred, we have also explored the relation between the CIR and evolution of the bars.

This paper is organized as follows. Section \ref{sec:section2} describes the properties of the sample galaxies followed by the data reduction techniques employed in this study. Section \ref{sec:section3} deals with various correlations while discussion and conclusion is provided in Section \ref{sec:section4}.

\section{THE DATA AND DATA REDUCTION}
\label{sec:section2}
The sample consists of 13 early-type spiral galaxies adopted from \citet{CM18} hosting nuclear rings. Their sample is taken from \citet{CO10} based on the primary criterion that the galaxies are observed by both \textsc{$HST$} and \textsc{$Spitzer$}. These galaxies have observations in at least four  \textsc{$HST$} bands. Also, the sample is devoid of galaxies possessing inclination $i >$ 70$^{\circ}$ and with central regions exhibiting dusty features \citep{CM18}. Most of the sample galaxies possess bars except NGC 7217, NGC 7742 and UGC 3789 which helped us  study the properties of the bars also. Though nuclear rings are preferentially found in barred spirals, a number of unbarred galaxies with other non-axisymmetric features such as strong spiral arms, were also reported to host nuclear rings as seen in NGC 7742 \citep{DE02}. We used archival \textsc{$HST$} images observed using $WFPC2/ACS$ instruments in \textsc{$F814W$} filter for our analysis. Though we used the standard pipeline calibrated images, the task \textit{L.A.Cosmic} \citep{VD01} and {\it{IRAF}} task {\it{cosmicrays}} were further used to improve the removal of cosmic rays. Out of the 17 galaxies used by \citet{CM18}, we  excluded galaxies with images that contained bad pixels in their central 3 arcsec region. We have excluded the galaxy NGC 7469 as the radius of the nuclear ring as estimated by \citet{CO10} is found to be within the 3 arcsec aperture.

\begin{table}
\centering
\begin{threeparttable}
\caption[]{The table lists the properties of sample galaxies: Name of the galaxy (column 1), Hubble Type (2), Distance modulus (3), non-axisymmetric torque parameter Q$_{g}$ (4), surface densities of ring clusters (5) average masses of the ring cluster population (6), average ages of the ring cluster population (7), relative size of the ring (8), the CIR computed in \textsc{$F814W$} band (9) and uncertainty in the estimation of the CIR (10). See table footnote, for the references for $a$ and $b$.}
\label{tab:table1}

\setlength{\tabcolsep}{1pt}
\small
 \begin{tabular}{llccccccccr}
  \hline\noalign{\smallskip}

		 Galaxy & Hubble type $^{[a]}$ & Distance modulus$^{[a]}$  & Q$_{g}^{[a]}$  & $\Sigma^{[a]}$  & log M$_{\rm cl}$/M$_{\odot}^{[a]}$ & log (t$_{\rm cl}$ yr$^{-1})^{[a]}$ & D$_{r}$/D$_{0}^{[b]}$ & CIR & $\Delta_{CIR}$ \\ 
		 &      &  (mag)           &                  & (kpc$^{-2}$)     &          &   &  &  &   \\
		 (1) & (2) & (3) & (4) & (5) & (6) & (7) & (8) & (9)  & (10) \\
  \hline\noalign{\smallskip}

ESO 565-11	&(R)SB(r)0/a&	34.23&	0.316&	8	&	5.56&	7.59 & 0.212&	1.06 &	0.08\\
NGC 1097&	SB(s)b&		31.4&	0.241&	213&	6.11&	8.83&	0.041&	0.58&	0.01 \\
NGC 1326&	(R)SB0+(r)&	30.86&	0.163&	484&		5.58&	8.11&0.048&	0.77&	0.03 \\
NGC 1512&	SB(r)a&		30.48&	0.366&	363&		5.15&	8.18&0.052&	0.72&	0.06 \\
NGC 1672 &	SB(s)b	&	30.81&	0.349&	3020&		6.25&	8.69&0.031&	0.56&	0.02 \\
NGC 2997&	SAB(rs)c&	30.2&	0.306&	179&	4.41&	7.72 &	0.016&	0.60&	0.04\\
NGC 3081 &	(R)SAB(r)0/a&	32.09&	0.194&	18&		6.39&	7.87&0.094&	0.88&	0.04 \\
NGC 4314 &	SB(rs)a	&	29.93&	0.432&	565&		4.6&	7.96&0.061&	0.74&	0.04 \\
NGC 6782 &	(R)SAB(r)a &	33.61 &	0.205&	7&		5.92&	8.59 &0.06&	1.41 &	0.06\\
NGC 6951&	SAB(rs)bc &	31.77&	0.275&	108&		6.08&	8.24&0.034&	0.61&	0.01 \\
NGC 7217&	(R)SA(r)ab&	31.41&	0.026&	49&		5.32&	8.64&0.082&	0.76&	0.04 \\
NGC 7742 &	SA(r)b	&	32.91  &	0.055 &	22 &		5.49&	7.44 &0.163&	0.96 &	0.05\\
UGC 3789 &	(R)SA(r)ab&	33.49&	-&	39 &	5.93&	7.52 &	-&	1.15 &0.03\\
  \noalign{\smallskip}\hline
\end{tabular}
\begin{tablenotes}\footnotesize
\item{Refs:~[a] \citet{CM18} and [b] \citet{CO10}}
\end{tablenotes}	
 \end{threeparttable}

\end{table}

\begin{figure}
\centering
	\includegraphics[scale=0.3]{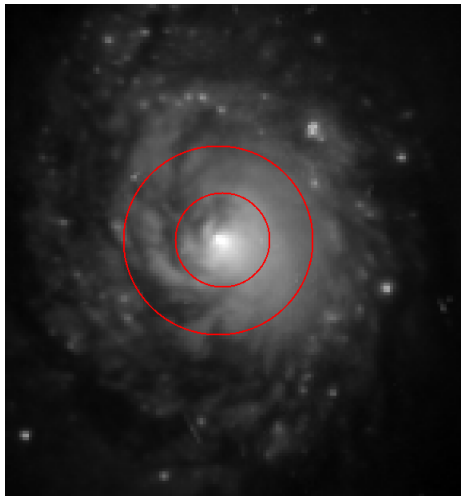}
    \caption{WFPC2 image of the galaxy NGC 1672 with the apertures used to calculate the CIR, shown in red colour.}
    \label{fig:figure0}
\end{figure}

The CIR for the sample galaxies  is determined using simple aperture photometry (MAG$\_$APER)  provided in source extractor (\textsc{$SEXTRACTOR$}, \citealt{BA96}). \citet{AR18} defined the CIR as,
\begin{equation}
   CIR = \frac{I_{1}}{I_{2} - I_{1}} = \frac{10^{0.4 (m_{2}-m_{1})}}{1-10^{0.4 (m_{2}-m_{1})}}.
	\label{eq:cir}
\end{equation}
where, $I_{1}$ (of magnitude $m_{1}$) and $I_{2}$ (of magnitude $m_{2}$) are the intensities of light within the inner and outer apertures of radii $r_{1}$ and $r_{2}$ (which is 2 $r_{1}$), respectively. The simple definition of CIR, helps avoid any dependence on a form following central intensity, $I (0)$ (i.e. surface brightness at a radial distance $r$, $I (r)$ = $I (0) f (r)$ where $f (r)$ is a function of $r$). On the other hand, the definition boosts any addition to (or subtraction from) the central intensity $I (0)$. Further, simple Monte Carlo simulations of ellipsoidal systems show remarkable stability over a range of radii (with $r_{2}$ = 2 $r_{1}$). However, in the present study, the inner radius is selected to contain the effects of the point spread function (PSF) in HST images while the outer radius is chosen to be far smaller than the half light radii of the sample galaxies. Thus, we have chosen the inner and outer radii as 1.5 and 3 arcsec, respectively, at a distance of 31.8 Mpc, which is roughly the mean distance of the sample galaxies. The physical distances corresponding to these radii are 0.23 and 0.46 kpc, respectively.  One of the typical images of our sample galaxies is shown in Figure \ref{fig:figure0} overlaid with the apertures used to calculate the CIR. The properties of the sample galaxies are provided in Table \ref{tab:table1}. The relative sizes of nuclear rings (see Section \ref{sec:section3.1}) are adopted from \citet{CO10} while other parameters are taken from \citet{CM18}. We have also listed the values of CIR  along with their uncertainties in Table \ref{tab:table1}.

\section{RESULTS}
\label{sec:section3}

We find that the CIR is closely associated with various properties of  nuclear rings. We also explore connections between the CIR and bar strengths of these galaxies.  The  linear correlation coefficients for all major correlations are given in Table \ref{tab:table2} along with the values of the best-fitting parameters.

\begin{table*}
\bc
\caption[]{The table lists the best-fitting parameters for the relation  x = $\alpha$ CIR + $\beta$.
\label{tab:table2}}
\setlength{\tabcolsep}{2.5pt}
\small
 \begin{tabular}{lcccccc}
  \hline\noalign{\smallskip}
x & $\alpha$ & $\beta$ & $r$ & $p$ & N \\
  \hline\noalign{\smallskip}
Q$_{\rm g}$  & -0.54 $\pm$ 0.17 & 0.63 $\pm$ 0.12 & -0.85 &  99.66 & 9 \\
        $\Sigma$ & -2.40 $\pm$ 0.58 & 3.97 $\pm$ 0.51 &  -0.78 &  99.82 & 12  \\
        D$_{r}$/D$_{0}$ & 0.35 $\pm$ 0.04 & -0.18 $\pm$ 0.03 & 0.94 & $>$ 99.99 & 11 \\
        M$_{cl}$/M$_{\odot}$ & 1.66 $\pm$ 0.45 & 3.82 $\pm$ 0.43 & 0.81 &  98.55 & 8  \\
        t$_{cl}$ & -2.13 $\pm$ 0.40 & 9.80 $\pm$ 0.33  &  -0.87 &  99.95 & 11\\ 
  \noalign{\smallskip}\hline
\end{tabular}
\ec
\tablecomments{0.98\textwidth}{N denotes the number of galaxies used in the fit. Pearson's linear correlation coefficient ($r$) is listed along with the significance ($p$).}
\end{table*}

\subsection{Correlations between the CIR and the Properties of Nuclear Rings}
\label{sec:section3.1}
Nuclear rings are believed to be closely associated with the activities in the central region  of host galaxies and its stellar population \citep{SA07,MA08,CM18}.  Evolution of a ring is found to depend on the dynamics of  the (associated) bar also \citep{BU99,CO10}. Yet, the photometric studies carried out so far seldom observed a direct connection between the properties of the ring and its host galaxy. 

\begin{figure} 
   \centering
   \parbox{6cm}{
\includegraphics[width=7cm]{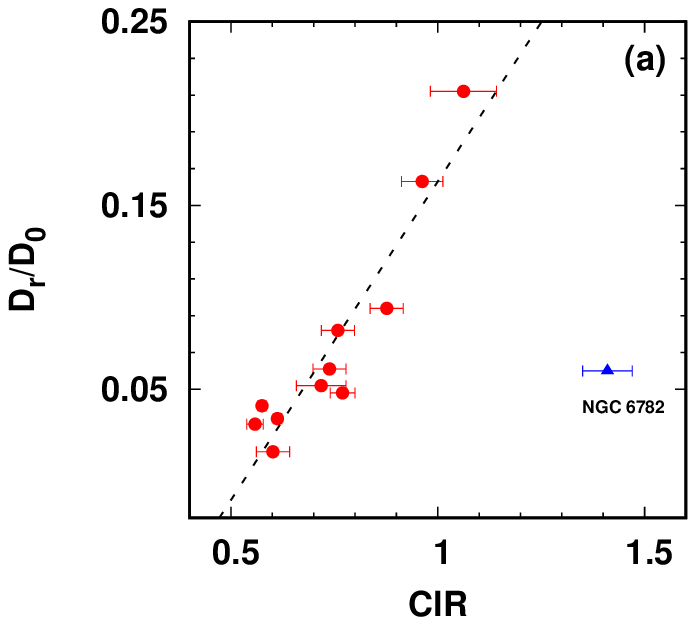}
\label{fig:figure1a}}
\qquad
\begin{minipage}{6cm}
\includegraphics[width=6.3cm]{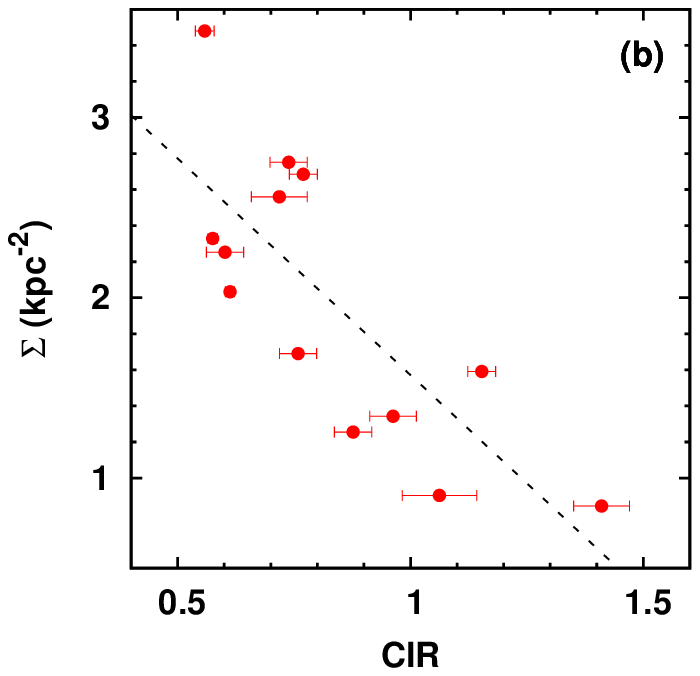}
\label{fig:figure1b}
\end{minipage} \\
\centering
\parbox{6cm}{
\includegraphics[width=7cm]{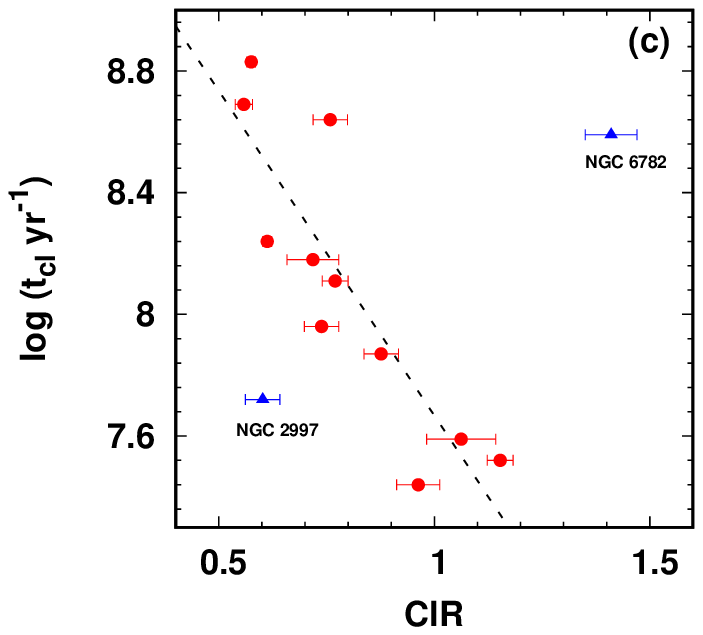}
\label{fig:figure1c}}
\qquad
\begin{minipage}{6cm}
\includegraphics[width=6.6cm]{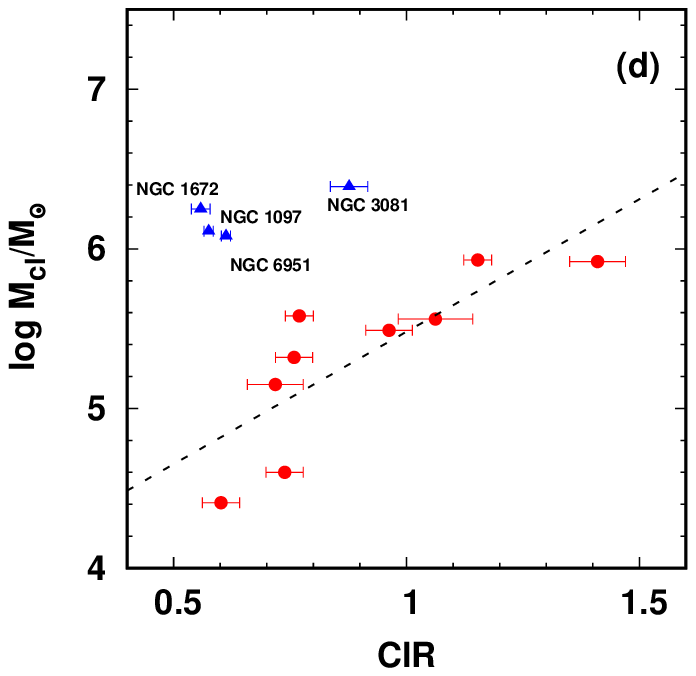}
\label{fig:figure1d}
\end{minipage} \\
   \caption{Correlations between the central intensity ratio and (a)  relative size of the ring, D$_{r}$/D$_{o}$, the ratio of the size of the ring to the diameter of its host galaxy adopted from \citet{CO10} (b) ring cluster surface density, $\Sigma$, (c) the average age of the ring cluster population, t$_{\rm cl}$ and (d) the average mass of the ring cluster population, M$_{\rm cl}$/M$_{\odot}$ taken from \citet{CM18}. The red coloured circles represent the galaxies following the fitted relations and outliers are denoted by blue triangles. } 
   \label{fig:figure1}
   \end{figure}

In the present study, we find a strong correlation between the CIR and the relative size of the ring (D$_{\rm r}$/D$_{\rm o}$), the size of the ring normalised with the diameter of the host galaxy, adopted from \citet{CO10}, as shown in Figure \ref{fig:figure1}(a). Here, D$_{\rm o}$ is the extinction-corrected D$_{25}$ radius of the galaxy as defined by \citet{BG95}. Pearson's linear correlation coefficient, $r$, is 0.94 with the significance, $P$ greater than 99.99 percent. We have used the recipe given in \citet{PT92} to calculate the significance of the correlations reported in this paper. This parameter is reported to be connected with the non-axisymmetric perturbation strength \citep{CO10}  and is further discussed in Section \ref{sec:section3.2}. The notable outlier which shows a large offset from the fitted relation in Figure \ref{fig:figure1}(a) is NGC 6782. This galaxy is seen to be an outlier in almost all correlations reported in this study. Further, we see that the ring cluster surface densities are anti-correlated with the values of CIR hinting at a close association of the CIR with the formation and evolution of the rings  (see Fig.\ref{fig:figure1}(b)). 

\citet{CM18} had performed detailed population synthesis of nuclear rings in the sample galaxies using Flexible Stellar Population Synthesis models \citep{CO09}. They fitted the observed spectral energy densities of the rings assuming a \citet{KR01} initial mass function (IMF) and solar metallicity to derive  average cluster ages and total stellar masses of the rings. The CIR shows a strong anti-correlation with the average cluster ages ($r=-0.87$, $P = 99.95\%$) while the correlation observed with the total stellar masses of the clusters is positive ($r = 0.81$, $P = 98.55\%$). The correlation between  the CIR and average age of the ring sample is shown in Figure \ref{fig:figure1}(c). The galaxies NGC 6782 and NGC 2997 are seen to be deviating from the fitted relation whereas in Figure \ref{fig:figure1}(d), we observe that all galaxies with masses above 10$^{6}$ M$_{\odot}$, deviate from the fitted relation.
   
\subsection {Correlation between the CIR and the Strength of the Bar}
\label{sec:section3.2}
\begin{figure}
\centering
	\includegraphics[scale=0.9]{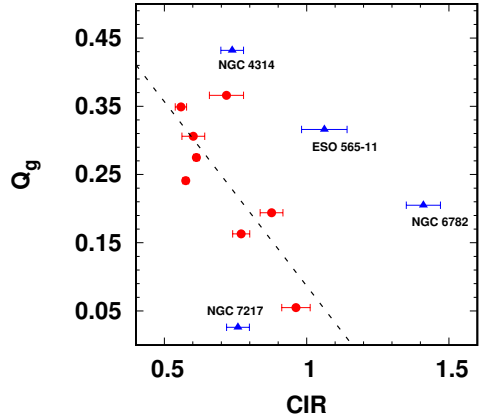}
    \caption{Correlation between the CIR and non-axisymmetric torque parameter (Q$_{g}$) adopted from \citet{CM18}. The symbols are similar to those used in Figure \ref{fig:figure1}.}
    \label{fig:figure2}
\end{figure}

The non-axisymmetric torque parameter (Q$_{g}$) was
first defined by \citet{CS81} as the ratio of the maximum tangential force to the mean radial force and it quantifies the strength of the bar. Higher values of Q$_{g}$ indicate stronger bars while lower values might be attributed to arm perturbations and oval distortions \citep{CO10}. In the case of galaxies without bars, this parameter may statistically be related to the strength of the spiral arms \citep{CM18}. We find a striking anti-correlation ($r$= -0.85 with $p$= 99.66 percent) between the CIR and Q$_{g}$ as shown in Figure \ref{fig:figure2}. Four galaxies (NGC 4314, NGC 7217, ESO 565-11 and NGC 6782) show significant offsets with respect to the fitted relation between the CIR and Q$_{g}$. The galaxy NGC 7217 is an unbarred galaxy. This galaxy is known to contain a counter rotating  stellar disc unlike the other galaxies in the sample \citep{CO10}. \citet{MA08} proposed that this galaxy might have under gone a minor merger in the past with a gas-rich dwarf galaxy. The galaxy ESO 565-11 hosts a highly elliptical ring which is currently being formed \citep{BU99}. This galaxy is also found to be an outlier in many of the correlations exhibited by galaxies hosting nuclear rings \citep{CO10}. Also, NGC 4314 is the youngest ring in the sample which is reported to be behaving differently \citep{CM18}. The  strong correlation between the bar strength and the CIR probably reflects the bar-driven star formation in the sample galaxies.

\section{DISCUSSION AND CONCLUSIONS}
\label{sec:section4}
We perform photometric studies of the centres of 13 early-type spirals (majority with bars) in the nearby Universe hosting nuclear rings. We use the CIR, a recently introduced measure of the concentration of light at the very centre of the galaxy image to study the formation and evolution of the nuclear rings and their host galaxies. The CIR is found to be a significant parameter in galaxy evolution studies as it is correlated with many structural and dynamical properties of early-type galaxies including the mass of the SMBH at the centre \citep{AR18}. However, estimation of CIR is not straight forward for spiral galaxies due to orientation effects posed by the disc component. But in the present study of spiral galaxies with nuclear rings, the orientation of most of the galaxies are face-on ($i <$ 70$^{\circ}$) enabling us to neglect the orientation effects induced by dust present in disc. Also, early-type spirals are known to be less heavily obscured by dust compared to late-type spirals \citep{CO10}. These features of our sample helped us unveil, for the first time, striking photometric correlations between the properties of nuclear rings and the CIR of their host galaxies. Also, we find that the bar strength of the sample galaxies is strongly correlated with the CIR reinstating the importance of the latter in galaxy evolution studies.

The present study indicates that the CIR is lower in strongly barred galaxies compared to the weak ones as quantified by the non-axisymmetric torque parameter. We know that the stellar bars in nearby spirals play a major role in the secular evolution of their host galaxies as they are capable of redistributing gaseous matter from the disks to inner regions of the galaxy \citep{SH76,AT92}. Such an infall of materials might trigger starbursts at the galactic centres \citep{HF97,HM99}. Several observational and theoretical studies  dealing with the bar-driven star formation in spirals suggest that this enhancement in star formation activity near the central region occurs at the onset of bar formation \citep[e.g.][]{HE80,EL11,FD15,SD17}. Once the bar is formed, the star formation starts declining and by the time the bar grows stronger, the star formation gets suppressed \citep{SD17,AA17}. 

The CIR contains information regarding the star formation activities near the central region of early-type galaxies \citep{AR18}. As seen in Figure \ref{fig:figure2}, the CIR decreases with the increasing strength of bars. The lower values of the CIR suggest that the star formation at the centre is minimum compared to the outer regions. This might be the result of the bar sweeping out the gaseous matter as it gets stronger and this is consistent with previous studies \citep{JP18}. Recently, \citet{CM18} investigated the star formation rates in nuclear rings of barred spirals and arrived at the conclusion that the SFR in rings is low in galaxies with strong bars. It is possible that the observed low values of CIR in such galaxies might be indicating low star formation in their very central region as well.

Also, we find that various properties of the nuclear rings correlate with the CIR. The smaller rings are found to possess lower values of the CIR as seen in Figure \ref{fig:figure1}(a). The ring cluster surface density is anti-correlated with the CIR,  re-affirming that in the galaxies with small and dense rings, the star formation activities at the centre is reduced. This might be a consequence of bar controlled evolution of nuclear rings (see for e.g. \citealt{KN05}). Many simulations also suggest that smaller rings occur in strongly barred galaxies \citep[e.g.][]{KS14}. \citet{KN05} propose that the nuclear rings can shrink as the bar gets stronger which seems to support  our observation. 

The average ring cluster masses and ages are also seen to be correlated with the CIR suggesting that old clusters which are less massive cause low values of central intensity ratio in their host galaxies. Except for four galaxies with masses above 10$^{6}$ M$_{\odot}$, the age of the ring clusters seem to be increasing with the CIR. It is tempting to propose an evolutionary history as follows: the clusters when they are formed due to starbursts are not only younger but also massive as they are rich in gaseous materials driven inwards by the newly formed bar. As the clusters get older, the bar gets stronger but the gas at the centre is exhausted due to the series of star bursts happening around the nuclear region. This results in a reduction of the intensity of light at the very centre of the galaxy as suggested by lower values of the CIR. Initially, when the starbursts occur, the light from the inner aperture is higher compared to the outer and therefore the CIR is higher.

The CIR is reported to be anti-correlated with the mass of the central SMBH in early-type galaxies. So far, studies have not been successful in unveiling any direct observational correlation between the active galactic nucleus (AGN) activities and occurrence of bars or formation of nuclear rings. The absence of intense star formation in the central region and strong AGN activities may be due to the prompt removal of gas in the central region \citep{HF97,HM99,KS00,LS02}. Recently, a numerical simulation exploring the AGN feedback and star formation history in barred spiral galaxies proposed that after the initial star burst, as a result of the inflowing gas materials towards the central AGN, the gas is gradually pushed away from the galactic centres shifting the star formation sites to larger radii \citep{RW17}. This scenario is also supported in the present study as we find low values of central intensity ratios in strong bars. In the case of early-type systems, it was observed that low values of central intensity ratio suggest massive black holes in galactic centres \citep{AR18}. Thus, the CIR  seems to possess the potential to play a crucial role in understanding the linked evolution of rings, bars and central AGNs. 

\normalem
\begin{acknowledgements}
We thank the anonymous reviewer for his/her valuable comments which greatly improved the contents of this paper. SA would like to acknowledge the financial support from Kerala State Council for Science, Technology and Environment (KSCSTE). We acknowledge the use of the NASA Extragalactic Database (NED), \href{https://ned.ipac.caltech.edu/}{https://ned.ipac.caltech.edu/} operated by the Jet Propulsion Laboratory, California Institute of Technology, and the Hyperleda database, \href {http://leda.univ-lyon1.fr/}{http://leda.univ-lyon1.fr/}. Some of the data presented in this paper were obtained from the Mikulski Archive for Space Telescopes (MAST), \href{http://archive.stsci.edu/}{http://archive.stsci.edu/}

\end{acknowledgements}
  
\bibliographystyle{raa}
\bibliography{bibtex}

\end{document}